\documentstyle[11pt,IAUS212,twoside,epsf]{article}

\def\edcomment#1{\iffalse\marginpar{\raggedright\sl#1\/}\else\relax\fi}
\reversemarginpar

\begin{document}
\vspace*{1cm}
\title{Outflows from ellipticals: the role of supernovae}
\author{Francesca Matteucci}
\affil{Department of Astronomy, University of Trieste, Via G.B. Tiepolo 11,
34100 Trieste, Italy}
\author{Antonio Pipino}
\affil{Department of Astronomy, University of Trieste, Via G.B. Tiepolo 11,
34100 Trieste, Italy}

\begin{abstract}
Models of SN driven galactic winds for ellipticals are presented. We assume 
that ellipticals formed at high redshift and suffered an intense burst of 
star formation. The role of supernovae of type II and Ia in the chemical 
enrichment and in triggering galactic winds is studied. In particular, 
several recipes for SN feed-back together with detailed nucleosynthesis 
prescriptions are 
considered. It is shown that SNe of type II have a dominant role in 
enriching the interstellar medium of elliptical galaxies whereas type Ia SNe 
dominate the enrichment and the energetics of the intracluster medium.
\end{abstract}

\section{Introduction}
Several mechanisms have been suggested so far 
for the formation and evolution of 
elliptical galaxies.
One scenario is based on an early monolithic collapse of a gas 
cloud or early merging
of lumps of gas where dissipation plays a fundamental role (Larson 1974;
Arimoto \& Yoshii 1987; Matteucci \& Tornamb\`e 1987). In this scenario
the star formation stops soon after a galactic wind 
develops and the galaxy evolves 
passively since then.
Bursts of star formation in merging subsystems made of gas
had been also suggested (Tinsley \& Larson 1979); 
in this picture star formation stops after the last burst and 
gas is lost via stripping or wind.
The alternative and more recent scenario is the so-called
hierarchical clustering scenario, where
merging of early formed stellar systems in a wide 
redshift range and preferentially at late epochs,
is expected (Kauffmann et al. 1993).
The main difference between the monolithic collapse and the 
hierarchical merging relies in the
time of galaxy formation, occurring quite early in the former scenario 
(at reshifts $z> 3$) and
continuously in the latter scenario.
There are arguments either in favour of the monolithic or
the hierarchical scenario, but the former one gives a more likely picture
since it can reproduce the majority of the properties of stellar populations 
in ellipticals, in particular some fundamental facts such as
that the ellipticals are dominated by old stars (K-giants) and 
that the [$\alpha$/Fe]$>0$ in the dominant stellar population
(Worthey et al. 1992; Weiss et al. 1995; Kuntschner et al. 2001). 
This high [$\alpha$/Fe] ratio is the clear signature of the pollution from 
massive stars. The same occurs in the most metal poor stars in our Galaxy 
and is due to the fact that SNe II are the main producers of 
$\alpha$-elements (O, Ne, Mg, Si. S and Ca) whereas SNe Ia, which 
explode with a delay relative to type II SNe, are thought to be 
responsible for the production of Fe.
Therefore, the high [$\alpha$/Fe] ratio in ellipticals argues strongly 
in favor of a short period of star formation during which type Ia SNe 
did not have time to substantially pollute the insterstellar medium (ISM).

In this paper we will discuss a monolithic model (Pipino et al. 2002) for 
the formation and evolution of ellipticals, where these objects suffer a 
short ($\le$ 1Gyr) but intense star formation period halted by a SN driven 
galactic wind. After the onset of the wind, which devoids the galaxy of 
all the gas present, star formation is assumed to stop. This is because 
the galaxy, after the wind, contains hot and 
rarified gas, a situation which is unfavorable to star formation. The time 
for 
the occurrence of the wind,
$t_{GW}$, is therefore crucial in determining the evolution of the galaxy 
and the intracluster/intergalactic medium.
Therefore, the assumptions about the energy transferred from SNe into the ISM
are very important. Unfortunately, very little is known about this 
feed-back and one has to choose the assumptions  
which produce a realistic model for ellipticals.
In section 2 we will discuss the condition for the occurrence of a 
galactic wind and the chemical evolution model. In section 3 we will 
present the results and draw a few conclusions.

\section{The model for ellipticals}
The model we are adopting here is described in detail in Pipino et al. (2002).
The main feature of the model is that it assumes a strong star formation rate
for ellipticals (roughly 20 times stronger than adopted in the solar 
vicinity) and takes into account in detail the contributions
from SN II and Ia.

\subsection{SN rates}
It is very important to compute detailed SN rates, taking into account 
the stellar lifetimes ($\tau_m$), in order to study the different roles 
played by SNII and SN Ia in galaxy evolution.

Type II SNe originates from the explosion of massive stars 
($M>10M_{\odot}$), the SNII rate is:
\begin{equation}
R_{SNII}=\,\int_{M_{m}}^{M_M}\psi(t-\tau_m) 
\phi(m)dm 
\end{equation}
with $M_m=10$ and $M_M=100M_{\odot}$ and $\phi(m)$ being the IMF.
Type Ia SNe originate from the thermonuclear explosion of
a CO-white dwarf (WD) in a binary system.
The binary system can be made of a CO WD plus a red giant star or by two 
CO WDs.
The type Ia SN rate in the single degenerate case can be written as:
\begin{eqnarray}
R_{SNIa}=A\int_{M_{Bm}}^{M_{BM}}
\phi(m)
\cdot[\int_{\mu_{min}}
^{0.5}f(\mu)\psi(t-\tau_{m2}) 
d\mu]dm\nonumber \\ 
\end{eqnarray}
where $\mu={M_2 \over M_B}$ (with $M_2$ mass of the secondary star) and 
$M_B$, the total mass of the binary system, 
is defined in the range
3-16 $M_{\odot}$ (see Matteucci \& Greggio, 1986).

\subsection{Stellar nucleosynthesis}
Type II SNe produce mainly $\alpha$-elements (O, Ne, Mg, Si, S, Ca)
and part of Fe. The adopted yields are from Woosley \&  Weaver (1995).
Type Ia SNe produce mainly Fe-peak elements ($\sim 0.6-0.7 M_{\odot}$ of Fe). The adopted yields are from Thielemann et al. (1993)
The yields from low and intermediate mass stars 
($0.8 \le M/M_{\odot} \le 8$) are from Renzini \& Voli (1981)

\section{The development of a galactic wind}
The condition for the occurrence of a wind, where for wind we intend an outflow which carries the gas out of the potential well of the galaxy, is:
\begin{equation}
(E_{th})_{ISM} \ge E_{Bgas}
\end{equation}
where $(E_{th})_{ISM}$ is the thermal energy of the gas and $E_{Bgas}$
is the potential energy of the gas.
\subsection{The thermal energy of gas}
The thermal energy of gas due to SN and stellar wind heating is:
\begin{equation}
(E_{th})_{ISM}=E_{th_{SN}}+ E_{th_{w}}
\end{equation}

with:
\begin{equation}
E_{th_{SN}}= \int^{t}_{0}{\epsilon_{SN}R_{SN}(t^{'})dt^{'}}
\end{equation}

and 
\begin{equation}
E_{th_{w}}=\int^{t}_{0}\int^{100}_{12}{ \varphi(m) \psi(t^{'}) \epsilon_{w}dm
dt^{'}}
\end{equation}
for the contribution from SNe and stellar winds, respectively.
The efficiencies of energy transfer from SNe into the ISM is
$\epsilon_{SN}= \eta_{SN}\epsilon_{o}$ 
with $\epsilon_o=10^{51}$erg 
(typical SN energy)
and that typical of stellar wind is:
$\epsilon_{w}= \eta_{w}E_{w}$ with $E_{w}= 10^{49}$erg 
(typical energy injected by a $20M_{\odot}$ star).

The simplest approach is to consider the $\eta_{SN}$ and $\eta_{w}$
efficiencies as constant.
Bradamante et al. (1998) estimated such efficiencies 
by computing the ratio between the energy 
in the shell of the supernova remnant and in the initial blast wave, 
at the time of the
merging of the shell with the ISM, by taking into account results from 
hydrodynamical calculations.
They found that under typical conditions, namely 
$\epsilon_{o}=10^{51}$ erg, $n_{o}=1cm^{-3}$ (density of the ISM)
and $c_{o}=10^{6}cmsec^{-1}$ (sound speed), 
$\eta_{SN}$=0.007-0.13 and $\eta_{w}\sim 0.03$.
Therefore, the majority of the initial blast wave energy 
of SNe and stellar winds is radiated away. However, this hypothesis is 
strictly valid for an isolated object and multiple SN explosions 
can radically 
change the situation. In addition, the contributions from different 
SN types can be different. For example,
SNe Ia exploding after type II should provide more energy 
into the ISM since they explode in a hot cavity.

In the present model we adopt a more complex formulation which assumes that 
the $\epsilon_{SN}$ is varying in time.
In particular, we assume the
formulation of Cox (1972) for the efficiency of energy injection from SNe:

\begin{equation}
\epsilon _{SN}=0.72 \epsilon_o \,\, erg
\end{equation}
for $t_{SN} \le t_c$ ,
where $t_c$ is the cooling time of a supernova remnant,
$\epsilon_o=10^{51}$ erg is the explosion energy and 
$t_{SN}$ is the time elapsed from the SN explosion.

For $t_{SN} > t_{c}$ holds:
\begin{equation}
\epsilon _{SN}= 2.2 \epsilon_o (t_{SN}/t_{c})^{-0.62}\,\, erg
\end{equation}
For the cooling time we adopt the formulation of
Cioffi et al. (1988) as 
a function of metallicity:
\begin{equation}
t_{c}=1.49 \cdot 10^{4} \epsilon_o^{3/14} n_{o}^{-4/7}(Z/ Z_{\odot})^{-5/14}\,\, yrs
\end{equation}

It is worth noting that also with these prescriptions $\eta_{SN}=0.01-0.02$.
Recchi et al. (2001) 
assumed that $\eta_{SNII}=0.03$ and $\eta_{SNIa}=1$ in successful
chemodynamical models
of dwarf irregular galaxies, in order to account for the fact that type 
Ia SNe occur with a delay and in an already hot cavity produced by 
type II SNe.
Pipino et al. (2002) tried several assumptions for SN feedback 
in ellipticals including this one and
concluded that no more than 35-40\% of the initial blast 
wave energy of SNe II+ Ia should heat the ISM 
in order to have realistic models for 
ellipticals and the intracluster medium (ICM).

\begin{figure}
\plottwo{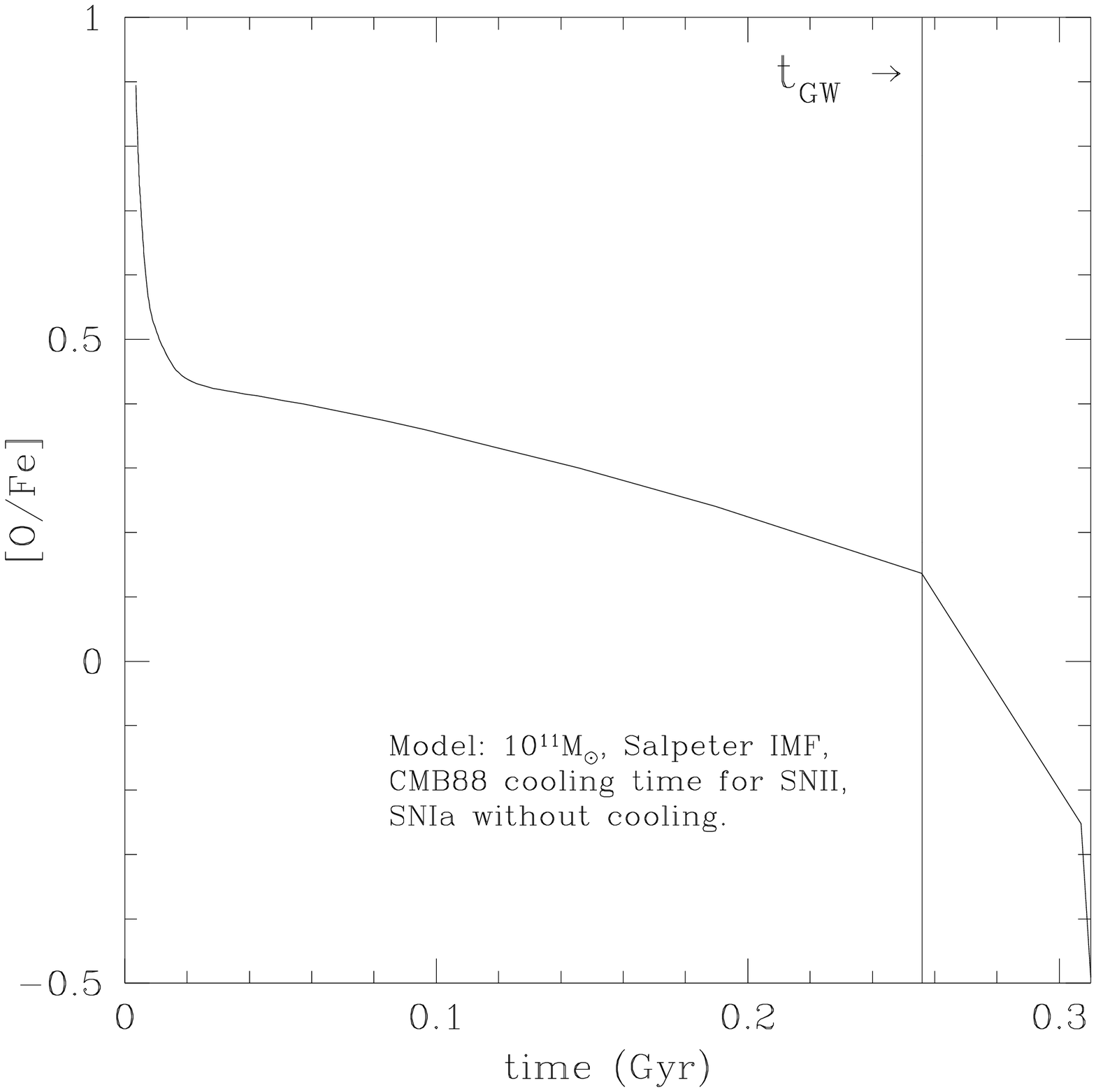}{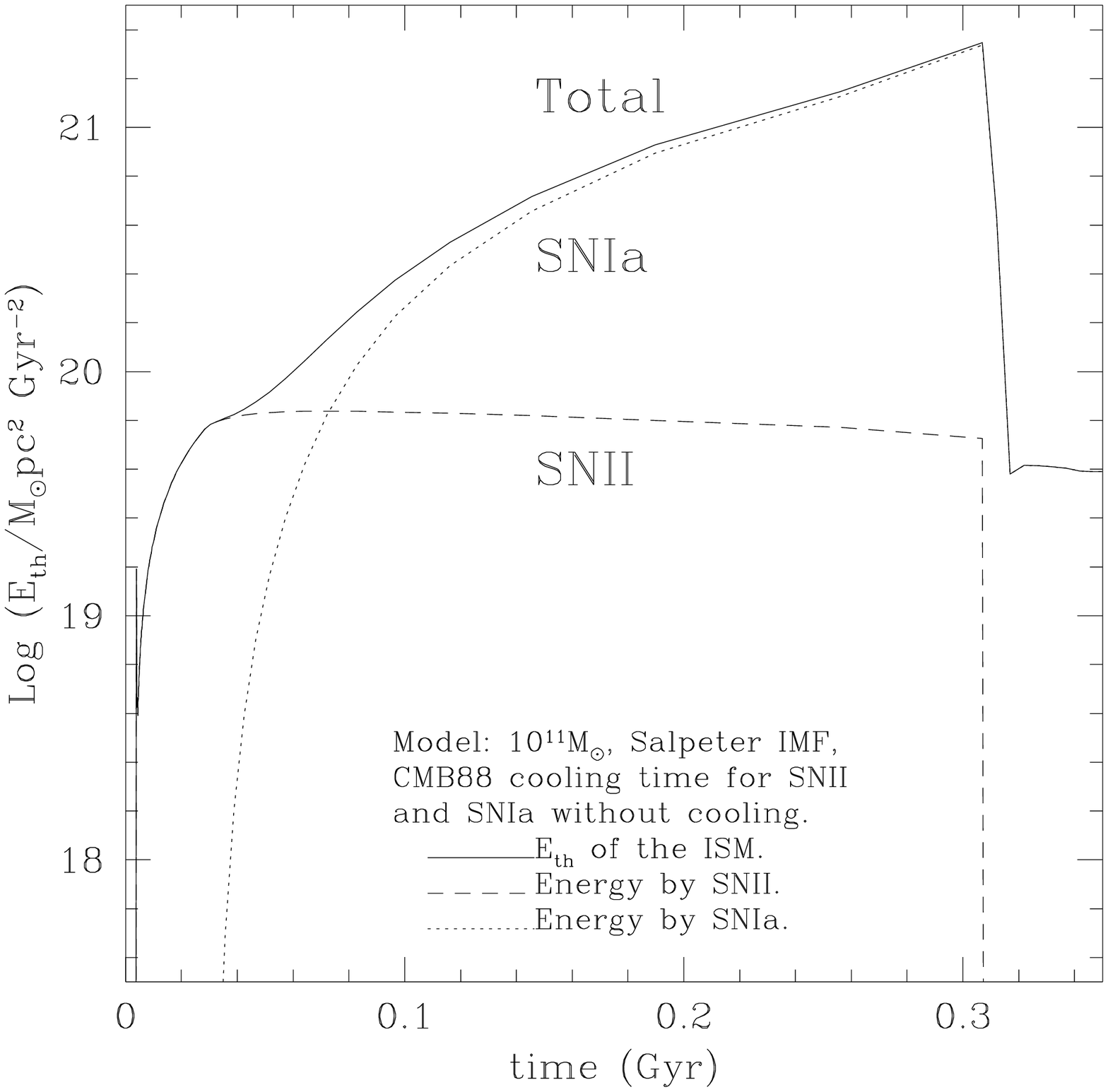}
\caption{Left figure: the predicted evolution of the [O/Fe] ratio as a function of time in the ISM
of an elliptical galaxy. The model assumptions (mass and feedback) are indicated in the figure. In particular CMB88 refers to Cioffi et a. (1988).
The time for the occurrence of the galactic wind is marked.
Right figure: predicted total thermal energy of the gas for the same galaxy of 
figure 1. The different contributions from SNII and Ia are indicated. 
SNeIa are favored energetically since we assumed $\eta_{SNIa}=1$. 
If $\eta_{SNII}=\eta_{SNIa}$ then SNII predominate in the energetics 
inside the galaxy
}
\end{figure}

\subsection{The potential energy of the gas} 

The total mass of the galaxy is expressed as $M_{tot}(t)=M_{*}(t)+M_{gas}(t)+M_{dark}(t)$
with $M_L(t)=M_{*}(t)+M_{gas}(t)$ being the luminous mass. The
binding energy of the gas is computed as in Matteucci (1992 and 
references therein).
For all galaxies here we assumed $M_{dark}/M_{L}=10$ and $r_L/r_{D}=0.1$, 
being the ratio between the effective radius and the radius of the dark 
matter core. 

\section{Model results}

The galactic winds in the models for ellipticals in the mass range 
$10^{9}-10^{11} M_{\odot}$ occur on a timescale less than 1 Gyr 
(Pipino et al. 2002).
Therefore, due to the short star formation period assumed for ellipticals, 
the 
abundance ratios in their stars show the signature of type II SN 
nucleosynthesis, namely
[$\alpha$/Fe]$>$0,  as it is evident in figure 1 where we show the 
predictions
of the model for the chemical composition of the gas in a galaxy 
with initial luminous mass $10^{11}M_{\odot}$, Salpeter (1955) IMF and 
feedback prescriptions as in Recchi et al. (2001).
In the figure is marked the age for the occurrence of a galactic wind
($t_{GW} \sim 0.25$ Gyr). After the wind no more star formation is 
taking place, 
so it is evident from the figure that the majority of the stars 
in this galaxy 
will show [$\alpha$/Fe] $>$ 0, whereas the gas that goes into the ICM will
show [$\alpha$/Fe]$\le 0$. This is due to the occurrence of 
type Ia SNe which have a maximum at 0.3-0.5 Gyr 
(Matteucci \& Recchi, 2001), 
namely after star formation has stopped.
The same is true for the energetics of type II and Ia SNe if we 
assume $\eta_{SNII}=\eta_{SNIa}$. 
On the other 
hand, if SNe Ia are assumed to inject all of their initial blast wave energy, 
the situation is reverted, as shown in figure 1 (right side); 
SNe Ia dominate the energetics even before 
the onset of the wind.This also implies that both the energetics and the 
chemistry of the ICM are dominated by type Ia SNe (see Pipino et al. 2002).
Finally, in figure 2 we show the predicted [X/Fe] vs. [Fe/H] relations for 
ellipticals of different initial luminous mass. Overimposed are the 
data for a Lyman-break galaxy MS 1512-cB58  at redshift $z=2.7276$
obtained by Pettini et al. (2002). 

\begin{figure}
\plotone{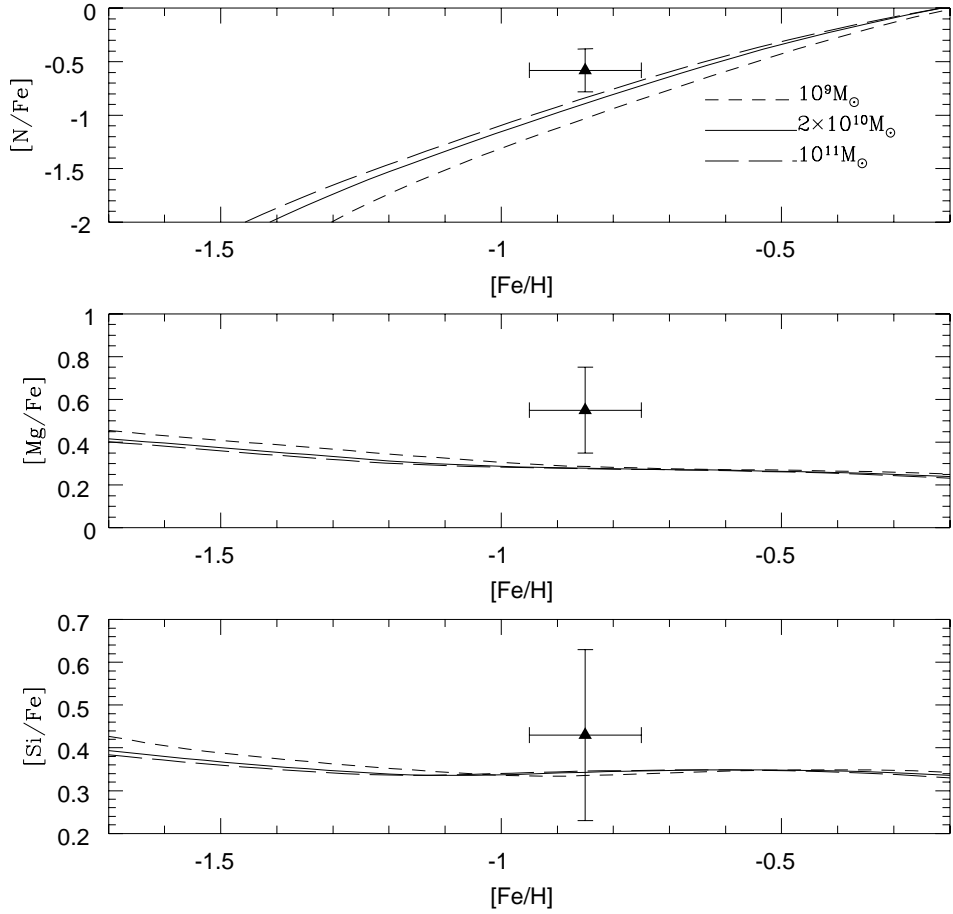}
\caption{Predicted [X/F] vs. [Fe/H] for models with different initial 
luminous masses, as indicated in the figure. The data point refers to the 
Lyman-break galaxy MS1512-cB58 from Pettini et al. (2002). The figure is from 
Matteucci and Pipino (2002).}
\end{figure}

The good agreement between the 
predictions for ellipticals and the observations strongly suggest that 
Lyman-break galaxies at high redshift could be ellipticals in formation, 
thus supporting the monolithic scenario.

\end{document}